# Pressure-induced superconductivity in a novel germanium allotrope

Liangzi Deng[1,#,*], Jianbo Zhang[2,#], Yuki Sakai[3], Zhongjia Tang[4], Moein Adnani[1], Rabin Dahal[1], Alexander P. Litvinchuk[1], James R. Chelikowsky[3,5], Marvin L. Cohen[6,7], Russell J. Hemley[8], Arnold Guloy[4], Yang Ding[2], Ching-Wu Chu[1,7,*]

\#: These authors contributed equally to this work

\* Corresponding: ldeng2@central.uh.edu, cwchu@uh.edu

1. Texas Center for Superconductivity and Department of Physics, University of Houston, Houston, Texas 77204, USA

2. Center for High Pressure Science and Technology Advanced Research, Beijing 100094, China

3. Center for Computational Materials, Oden Institute of Computational Engineering and Sciences, University of Texas at Austin, Austin, Texas 78712, USA

4. Texas Center for Superconductivity and Department of Chemistry, University of Houston, Houston, Texas 77204, USA

5. Department of Physics and McKetta Department of Chemical Engineering, University of Texas at Austin, Austin, Texas 78712, USA

6. Department of Physics, University of California at Berkeley, Berkeley, California 94720, USA

7. Materials Sciences Division, Lawrence Berkeley National Laboratory, Berkeley, California 94720, USA

8. Departments of Physics, Chemistry, and Earth and Environmental Sciences, University of Illinois Chicago, Chicago, Illinois 60607, USA

**Abstract**

High-pressure studies on elements play an essential role in superconductivity research, with implications for both fundamental science and applications. Here we report the experimental discovery of surprisingly low pressure driving a novel germanium allotrope into a superconducting state in comparison to that for α-Ge. Raman measurements revealed structural phase transitions and possible electronic topological transitions under pressure up to 58 GPa. Based on pressure-dependent resistivity measurements, superconductivity was induced above 2 GPa and the maximum $T_c$ of 6.8 K was observed under 4.6 GPa. Interestingly, a superconductivity enhancement was discovered during decompression, indicating the possibility of maintaining pressure-induced superconductivity at ambient pressure with better superconducting performance. Density functional theory analysis further suggested that the electronic structure of Ge (oP32) is sensitive to its detailed geometry and revealed that disorder in the β-tin structure leads to a higher $T_c$ in comparison to the perfect β-tin Ge.

Keywords: Superconductivity; Element superconductor; High Pressure; Phase transition; Allotrope

Highlights:

- Superconductivity emerges in Ge (oP32) at low pressure.
- A decompression-driven superconductivity enhancement is discovered.

- A pressure-driven electronic topological transition favorable to superconductivity is suggested.

1. Introduction

The high-pressure behavior of Group 14 elements has attracted great attention due to its unresolved fundamental scientific questions, as well as its significant technological applications. Upon compression, the semiconducting diamond structure of Ge undergoes several structural transformations, entering the metallic β-tin phase with the space group $I4_1/amd$ at ~ 10 GPa [1], the *Imma* phase at ~ 75 GPa [2], the simple hexagonal structure with the space group $P6/mmm$ at ~ 85 GPa [3], the orthorhombic *Cmca* phase near 100 GPa [4], and the hexagonal close packed (hcp) structure near 160-180 GPa [4]. Superconductivity emerges above 10 GPa as the crystal transforms into the β-tin phase with a superconducting transition temperature ($T_c$) ~ 5.4 K, followed by a negative $dT_c/dP$ with further increasing pressure [5]. It was suggested that the standard electron-phonon coupling mechanism is responsible for the superconductivity in Ge [6]. One of the aims of the current study is to investigate the possibility of lowering the required pressure for inducing superconductivity in Ge. Allotropes of an element can display drastically different properties depending on the fundamental chemical nature of the element. A metastable germanium allotrope, oP32-structured Ge, was successfully synthesized and was found to be a diamagnetic semiconductor at ambient pressure [7]. Interestingly, we discovered that the Ge (oP32) single crystal becomes superconducting with a $T_c$ ~5.5 K above 2 GPa, which is much lower than the pressure required for α-Ge to turn into a superconductor and warrants further low-temperature structural studies. It was suggested that the β-tin phase could exhibit a high $T_c$ if it were formed under a lower pressure [8]. In addition, a rare decompression-induced enhancement of $T_c$, similar to phenomena reported in other systems [9], was also observed.

2. Methods

2.1 Sample Preparation

Ge (oP32) single crystals were synthesized based on the procedures described in our previous study [7]. The precursor $Li_7Ge_{12}$ was prepared by placing stoichiometric amounts of pure elements (Li ribbons and Ge pieces) in Ta tubes and sealing the tubes by welding under an argon atmosphere, followed by further enclosing them in sealed vacuum quartz tubes. The reaction charges were heated to 940 °C for 1 hour, after which each product was reground and subsequently annealed at 490 °C for 7 days. The pure $Li_7Ge_{12}$ powder samples were then each placed into a glass-bottom container with a reacted ionic liquid [dodecyltrimethylammonium aluminum tetrachloride (DTAC) or hexyltrimethylammonium aluminum tetrabromide (HTMAB)], and the glass-bottom container was further enclosed in a closed glass jacket and heated to 135-140 °C for 3-7 days. The dark grayish solid products were separated from the reaction mixture by centrifugation. By using acetone, shiny, relatively large crystals were separated out from the products. These crystals were subsequently determined to be Ge (oP32) by single-crystal powder X-ray diffraction (XRD) and no traces of α-Ge were detected (Fig. S1). Powder samples from both DTAC and HTMAB reactions resulted in a single phase (> 95% by XRD) of the allotrope. The allotrope is relatively stable to air and moisture.

2.2 Raman Measurements under Pressure

*In situ* high-pressure Raman scattering measurements were performed on single crystals of Ge (oP32) by using a symmetric diamond anvil cell (DAC) with 300 μm culet size diamond anvils and collecting the scattered light in the backscattering geometry from 300 K to 7 K and under magnetic fields up to 5 T. Ne was loaded inside as the pressure transmitting medium (PTM) and ruby balls were used for pressure measurement. Pre-pressed rhenium with a hole 100 μm in diameter was used as the gasket to hold the sample under high pressure. The 633 nm line from a He–Ne laser was used for excitation with power of less than 1 mW. The 1200 lines/mm grating enables spectral resolution of 0.45 cm$^{-1}$.

## 2.3 Electrical Transport Measurements under Pressure

For resistivity measurements conducted in this investigation, the pressure was applied to the samples using a symmetric DAC. We used anvils with a 400 μm culet. The rhenium gasket was insulated with Stycast 2850FT. The sample's chamber diameter was ~ 170 μm, where cubic boron nitride (cBN) was used as the PTM. Samples were cleaved and cut into thin squares with a diagonal of ~ 150 μm and thickness of ~ 20 μm. The pressure was determined using the ruby fluorescence scale [10] or the diamond Raman scale at room temperature [11]. The samples' contacts were arranged in a van der Pauw configuration and data were collected using a Quantum Design Physical Property Measurement System (PPMS) with temperatures down to 1.9 K and magnetic fields up to 7 T.

## 2.4 Theoretical Calculations

We employed a total-energy pseudopotential method constructed within density functional theory [12-15] to study the electronic structure of Ge (oP32) under pressure. Troullier-Martins norm-conserving pseudopotentials combined with a plane-wave basis were used for electronic structure simulations [16,17]. Exchange and correlation functionals are approximated within the local density approximation. A *4 × 4 × 4 k*-grid and a cutoff energy of 50 Ry are sufficient to obtain converged electronic properties. Variable-cell relaxation was conducted to obtain structures under hydrostatic pressure. We also performed variable-cell molecular dynamics simulations under constant pressure based on the method proposed by Parrinello and Rahman [18]. We used density functional perturbation theory to simulate vibrational and superconducting properties of Ge-based materials [19,20]. A Gaussian broadening of 0.01-0.02 Ry is used to approximate delta functions, which appear in Brillouin zone summations of electron-phonon calculations. We computed the Eliashberg spectral function and electron-phonon coupling strength λ to estimate the superconducting transition temperature $T_c$ by using the Allen-Dynes equation [21]. Here we used an effective Coulomb repulsion parameter μ* of 0.12.

## 3. Results and Discussion

### 3.1 Raman spectra and phase transitions under high pressure

The orthorhombic crystal structure of Ge (oP32) is shown in Fig. 1a. The coordination around each Ge atom is distorted from tetrahedral symmetry with bond angles ranging from 93° to 127°. The layered structure is demonstrated in Fig. 1b. Ge (oP32) was reported to be a semiconductor with an energy gap $E_g$ = 0.33 eV [7]. At ambient pressure, preliminary magnetic measurements show diamagnetic behavior and an irreversible transformation of Ge (oP32) to α-Ge was observed at 636 K [7].

Systematic room-temperature Raman measurements were carried out under different pressures. Based on site-symmetry group theoretical analysis, 48 Raman-active modes are expected as $\Gamma_{Raman} = 13A_{1g} + 14B_{1g} + 11B_{2g} + 10B_{3g}$ [22], and 27 have been observed experimentally [7, 23]. As shown in Fig. 2, the 19 modes we observed are close to those previously reported [7, 23]. The absence of the others is likely due to their low scattering intensity or the overlap of closely lying bands. For each mode at a given symmetry, one could easily find a counterpart among the modes calculated using density functional theory (DFT) [23]. It is important to note that we observed the unique mode with frequency 316.1 cm$^{-1}$ of Ge (oP32), which corresponds to the stretching vibration of the shortest interlayer (Ge1−Ge3) bonds [2.4052(15) Å] reported previously [7, 23]. In contrast, no characteristic frequency above 300 cm$^{-1}$ has been found in α-Ge [24, 25], hexagonal 4H-Ge [24, 26], or microcrystalline Ge allotrope (m-allo-Ge) [24] either experimentally or from DFT calculations. We also observed one Raman vibration mode at 87.6 cm$^{-1}$, theoretically derived from $B_{3g} + B_{1g}$, that has not been reported previously. In addition, the sharpness of the observed phonon lines even at room temperature is evident from the experimental Raman spectra of Ge (oP32) single crystals, as shown in Fig. 2a. This indicates unambiguously the high crystallinity of the Ge (oP32) allotrope.

Two phase transitions are shown in Fig. 2a and 2b, as indicated by different colors and by dashed vertical lines, respectively. Phase I (below 9.8 GPa) is orthorhombic, in which the softening of modes ≤ 100 cm$^{-1}$ and the hardening of modes > 160 cm$^{-1}$ was observed. Three specific observations are worth mentioning: 1) the peak at 166.2 cm$^{-1}$, which has not been reported previously, appears at 0.2 GPa and disappears at 7.5 GPa; 2) the two peaks at 222.8 cm$^{-1}$ and 267.3 cm$^{-1}$, corresponding to the $B_{3g} + B_{1g}$ and $A_{1g}$ modes, respectively, merge with their respective adjacent peaks around 1.4 GPa; and 3) the peak at 185.4 cm$^{-1}$ corresponds to the $A_{1g}$ mode and changes little with increasing pressure as shown by the evolution of the vibration modes as a function of pressure displayed in Fig. 2b. These phenomena indicate the existence of electronic topological transitions in Phase I. Phase II (9.8 GPa – 15.4 GPa) is a mixed phase, in which the structure gradually changes from orthorhombic to β-tin. Some of the vibration modes from Phase I remain. The peak at 316 cm$^{-1}$, characteristic of Ge (oP32), disappears in Phase II, and five new modes (89.1, 205.7, 252.7, 268.7, and 275.8 cm$^{-1}$), appear at 9.8 GPa. A comparison with the Raman-active modes in the BC8 structure [87 cm$^{-1}$ (Tg), 214 cm$^{-1}$ (Tg), 227 cm$^{-1}$ (Ag), 246 cm$^{-1}$ (Tg), and 259 cm$^{-1}$ (Eg) at 0 GPa] indicates that Phase II may contain part of the BC8 structure. Another two new Raman vibration modes that appear at 81.1 cm$^{-1}$ under 13.3 GPa and at 218.9 cm$^{-1}$ under 9.8 GPa are considered to be characteristic of β-tin [27]. In Phase III (above 15.4 GPa), the high-pressure β-tin phase survives up to 57.3 GPa, the highest pressure we applied during the Raman measurements. As shown in Fig. 2b, the higher frequency Phase III peak is more sensitive to pressure change than that associated with the lower frequency, which is consistent with the results of our DFT lattice dynamics calculations and earlier reported Raman spectra evolution of β-tin under high pressure [27].

Additionally, as shown by the Raman spectra displayed in Fig. S2, Ge (oP32) remained stable when the temperature was decreased to 7 K and the magnetic field was increased to 5 T. Moreover, the observed phonon lines become sharper at low temperatures, indicating unambiguously the high crystallinity of the Ge (oP32) allotrope. We also carried out high-pressure Raman measurements at 100 K and observed two phase transitions (Fig. S3), which further confirmed our experimental results at room temperature and high pressure.

Many studies have shown that, upon decompression, Ge does not follow the reverse of the structural sequence observed during compression. Depending on the pressure release process, metastable back-transformations lead to new allotropes, *i.e.*, the BC8 structure gradually changes to the hexagonal diamond structure [28], the metastable ST12 structure [29], and the R8 structure [30] at different decompression rates, and hydrostaticity may cause the varying phase selection [31]. Therefore, we collected Raman spectra of Ge (oP32) during decompression to get a better understanding of the phases retained following different pressure cycles.

We obtained Raman spectra for Ge (oP32) during decompression and found even more intriguing results. Figure 3a displays the Raman spectra following the release of pressure after the sample was pressurized to 57 GPa, with all compression and decompression processes performed at room temperature. Phase III was retained at 10.6 GPa. When the pressure was further decreased to 6.1 GPa, Phase II was restored. The critical pressure for each phase transition during decompression is different from that during pressurization. Similar hysteresis behavior was also observed in amorphous Ge [32] and amorphous Si [33, 34] at slightly higher pressure. Interestingly, the intensity of the Raman spectra becomes weaker when the pressure is further reduced to 0.4 GPa. However, it is clear that it does not return to the initial state, as shown in Fig. 3c, although we were not able to determine the resulting structure. To confirm that the difference between the release pressure of 0.4 GPa and the initial state is not due to sample damage or degradation or to high-pressure-induced weak links, we performed Raman measurements during decompression in a lower pressure range, and the obtained Raman spectra are shown in Fig. 3b. We first released the pressure from 11 GPa to 6.0 GPa and the resulting spectrum is fundamentally consistent with that obtained at 6.1 GPa after decompression from higher pressure, as shown in Fig. 3a. We continued to release the pressure to 1.0 and 0.6 GPa and found that the sample does not return to Phase I. We then uninstalled the diamond anvil cell (DAC) and obtained the Raman signal at ambient pressure. It is clearly stronger without diamond attenuation, but, as compared to the spectrum shown in Fig. 3c, it also does not return to the initial state. By comparing our results with those in a previous report [35], we confirmed that the observed phase is the ST12 phase. It is possible that the hexagonal diamond (lonsdaleite) and R8 phases [36] were formed when the sample was decompressed to 1.0 or 0.6 GPa, as shown in Fig. S4. Based on both Figs. 3a and 3b, it can be realized that Ge (oP32) is a metastable metal below 6.0 GPa during the decompression process, and its metastable structure has a strong dependence on the specific process employed. Similar behavior has been observed in other types of Ge allotropes [32]. It is possible that the BC8 phase was formed when the sample here was decompressed from the β-tin phase [28], which could be the cause of the abnormal $T_c$ increase discussed below [37]. Given the close relationship between its structure and physical properties, the physical properties of Ge (oP32) are expected to vary during pressurization and decompression.

### 3.2 Pressure-induced superconductivity and enhancement during decompression

With increasing pressure, the signature of superconductivity appears in Ge (oP32) at 2.9 GPa with an onset $T_c \sim 5.7$ K (Fig. 4). The $T_c$ reaches a maximum at ~ 6.8 K under 4.6 GPa, followed by a $dT_c/dP \sim -0.1$ K/GPa with further increasing pressure. Zero resistance is observed at 12.4 GPa, as shown in Fig. S5. Although the resistance does not reach zero when the transition kicks in at lower pressure, the evolution of this transition under pressure and the magnetic field effect on the transition temperature (Fig. S7) are consistent with superconducting behavior. A clear decompression-induced superconductivity enhancement is observed between 10 GPa and 30 GPa, as demonstrated in Fig. S6. Interestingly, no superconductivity signature was detected down to 1.9 K below 9 GPa during decompression, as indicated by the gray rectangle in Fig. 4. The magnetic field *vs.* temperature phase diagram for Ge (oP32) at 5.7 GPa and 11.9 GPa is shown in Fig. S7. Based on the Ginzburg-Landau (GL) fitting [38], $H_{c2}$ is ~ 2.4 T under 5.7 GPa and ~ 4.9 T under 11.9 GPa.

### 3.3 Band structure and phonon dispersion relation under high pressure

Density functional theory (DFT) studies were carried out to further understand the mechanism underlying the low pressure needed to induce superconductivity in Ge initially in the oP32 phase. Figure 5a shows the band structure of Ge (oP32) under pressure. At ambient pressure, Ge (oP32) shows a direct gap of 0.27 eV at the Γ point. Under 5 GPa, it becomes an indirect band gap between the Γ and X points and the gap is increased to 0.54 eV. The indirect gap decreases to 0.28 eV at 10 GPa due to the lowering of the conduction band around the X point. The structural changes are not strong under these pressures (see Fig. 5b). Under pressure of 5 or 10 GPa, some germanium atoms are shifted slightly from their original positions at 0 GPa, but the overall structure is not substantially different. This indicates that the electronic structure of oP32 is sensitive to its detailed geometry. Raman measurements found no significant structural phase transition below 10 GPa. Therefore, the change of the energy band structure indicates that the Fermi surface topology changes, and a possible pressure-induced Lifshitz transition occurs. At even higher pressure, the oP32 structure undergoes an insulator-to-metal transition. Figure 5c shows the band structure of oP32 under 12.5 GPa. In this case, hybridization of both the valence and conduction bands is strong and several bands cross the Fermi energy, indicating a complicated Fermi surface. The structure is also strongly modified despite the small jump in pressure from 10 GPa to 12.5 GPa. Due to this strong modification, some atoms are close to being 6-fold coordinated, as shown in Fig. 5d, similar to that in the β-tin structure. Such 6-fold coordination should be the origin of the observed insulator-to-metal transition.

The phonon dispersion relation of Ge (oP32) under various pressures is presented in Fig. 6. No signs of instability are found in the phonon dispersion under pressures ≤ 10 GPa. Following the insulator-to-metal transition at 12.5 GPa, the phonon dispersion relation of Ge (oP32) shows an imaginary frequency mode around the X point, indicating the instability of this structure. If we neglect the imaginary phonon frequencies, the $T_c$ is about 1.6 K at 12.5 GPa. This is lower than that of β-tin Ge around the same pressure range. From a screenshot structure of variable cell molecular dynamics (MD) around 4 ps at 12.5 GPa, we obtained the β-tin structure after full structural relaxation. A similar procedure at 15 GPa yields a disordered β-tin with imaginary frequency phonon branches. Interestingly this distorted β-tin exhibits a higher $T_c$ (9.2 K) than β-tin (4.1 K at 15 GPa), probably because of the disorder, as reported previously [39]. These results indicate that the high-pressure structure is β-tin-like, but some parts could be disordered since the transition from Ge (oP32) to β-tin is not simple compared to that from diamond to β-tin. The Ge (oP32) structure is relatively sensitive to pressure and the volume reduction is faster compared to that of the diamond-structure germanium. When comparing results at 0 and 10 GPa, the Ge (oP32) volume is reduced by 14%, while the volume reduction in the diamond structure is 11%. Further details regarding the superconducting parameters and crystal structures are summarized in Table S1 and Table S2.

The difference between the critical transition pressure (from the non-superconducting state to the superconducting state) observed experimentally and that determined theoretically could be due to the following factors. 1) We mainly used a variable-cell relaxation method to obtain the structures, which may be too gentle to determine the optimized structure. 2) More flexibility in Ge (oP32) causes an experimental structural phase transition or an electronic topological transition [40] even under a relatively low pressure of 2-3 GPa. Further structural investigation under high pressure and low temperature is critical to understand the mechanism for the superconductivity induced in Ge (oP32) at low pressure. In addition, it would be worth investigating approaches to further lower the pressure required to retain the pressure-induced superconductivity in this system, as demonstrated in our recent studies [41, 42].

Based on our calculation results and experimental data, it is safe to conclude that the Ge (oP32) crystal begins to turn into β-tin Ge above 10 GPa. It is worth mentioning that the Ge crystal will transform into the tetragonal structure (β-tin Ge) under ~10 GPa whether it starts as a cubic structure (α-Ge) or as an orthorhombic structure [Ge (oP32)]. The former case is a phase transition to lower symmetry, while the latter is a pressure-driven phase transition to higher symmetry. However, regarding the superconductivity in this crystal at the lower pressure range, *i.e.*, 3-10 GPa, there may be a few possibilities: 1) it is still in the Ge (oP32) phase but becomes superconducting; 2) the β-tin phase is formed at low pressure [8] with a $T_c$ higher than ever previously observed; and 3) a new phase, neither Ge (oP32) nor β-tin Ge, forms.

4. Conclusion

The present study unveils superconductivity induced under a much lower pressure in Ge (oP32) compared to α-Ge, which is possibly related to a pressure-driven electronic topological transition. We suggest that the simulation results provided here cannot reproduce the $T_c$ experimentally observed below 10 GPa due to the difficulty in determining the sample structure in the high-pressure, low-temperature superconducting state. We cannot entirely exclude the possibility that the standard simulation based on density functional perturbation theory [31, 32] and the Allen-Dynes equation [33] does not correctly describe the superconductivity in Ge (oP32). A decompression-driven superconductivity enhancement, which might be due to the formation of a mixed phase including the BC8 structure, was also observed. Our findings open new avenues for expanding the scope of superconductors with notably lower pressure barriers that are more adaptive and suitable for applications in diverse and demanding implementation environments.

## ACKNOWLEDGMENTS

L.Z.D., M.A., R.D., and C.W.C acknowledge support by U.S. Air Force Office of Scientific Research Grants FA9550-15-1-0236 and FA9550-20-1-0068, the T. L. L. Temple Foundation, the John J. and Rebecca Moores Endowment, the State of Texas through the Texas Center for Superconductivity at the University of Houston (TcSUH). J.B.Z and Y.D. acknowledge support by the National Key Research and Development Program of China (Grant Nos. 2018YFA0305703 and 2022YFA1402301) and the National Natural Science Foundation of China (Grant Nos. U1930401 and 11874075). Z.J.T. and A.G. acknowledge support by the Robert A. Welch Foundation E-1297 and the State of Texas through TcSUH. Y.S. and J.R.C. acknowledge support by a sub-award from the Center for Computational Study of Excited-State Phenomena in Energy Materials (C2SEPEM) at the Lawrence Berkeley National Laboratory, which is funded by the U.S. Department of Energy under contract no. DE-AC02-05CH11231, as part of the Computational Materials Sciences Program. J.R.C. also acknowledges support from the Welch Foundation under grant F-2094. The National Energy Research Scientific Computing Center and the Texas Advanced Computing Center provided computational resources. M.L.C. acknowledges the support of National Science Foundation (NSF) Grant No.DMR-2375410. R.J.H. acknowledges the NSF Grant DMR-2104881 and Chicago/DOE Alliance Center (CDAC) DOE-NNSA cooperative agreement DE-NA0003975.

**Author Contributions**

Liangzi Deng: Conceptualization, Methodology, Investigation, Writing – original draft, Writing – review & editing, Supervision. Jianbo Zhang: Methodology, Investigation, Writing – original draft, Writing – review & editing; Yuki Sakai: Methodology, Writing – original draft, Writing – review & editing; Zhongjia Tang: Investigation; Moein Adnani: Investigation; Rabin Dahal: Investigation; Alexander P. Litvinchuk: Investigation; James R. Chelikowsky: Supervision; Mavin L. Cohen: Supervision; Russell J. Hemley: Supervision; Arnold Guloy: Supervision; Yang Ding: Supervision; Ching-Wu Chu: Writing – original draft, Supervision.

**Competing Interests.** The authors declare no competing interests.

**Data Availability.** The authors declare that the data supporting the findings of this study are available within the paper and its Supplementary Information.

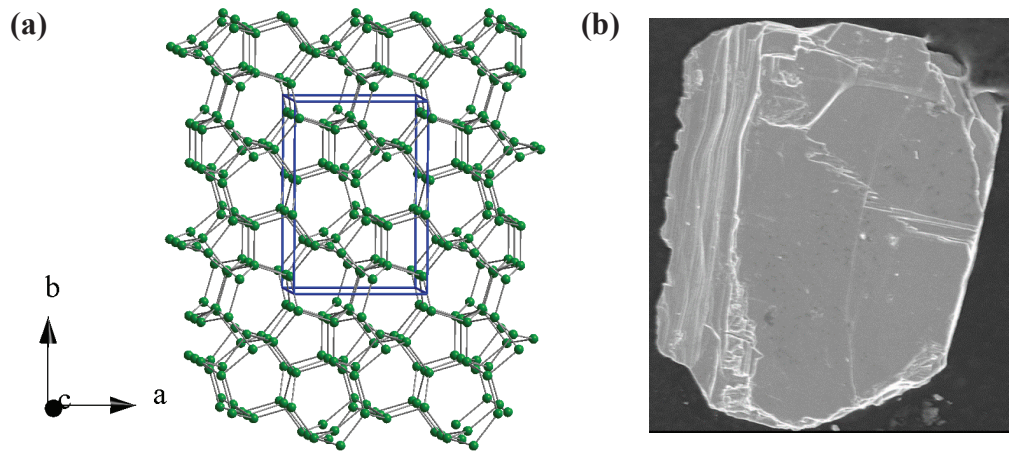

Fig. 1 (a) Crystal structure of Ge (oP32). (b) Scanning electron microscopy (SEM) image of a Ge (oP32) single crystal showing its layered structure. The crystal plane is along the [010] direction.

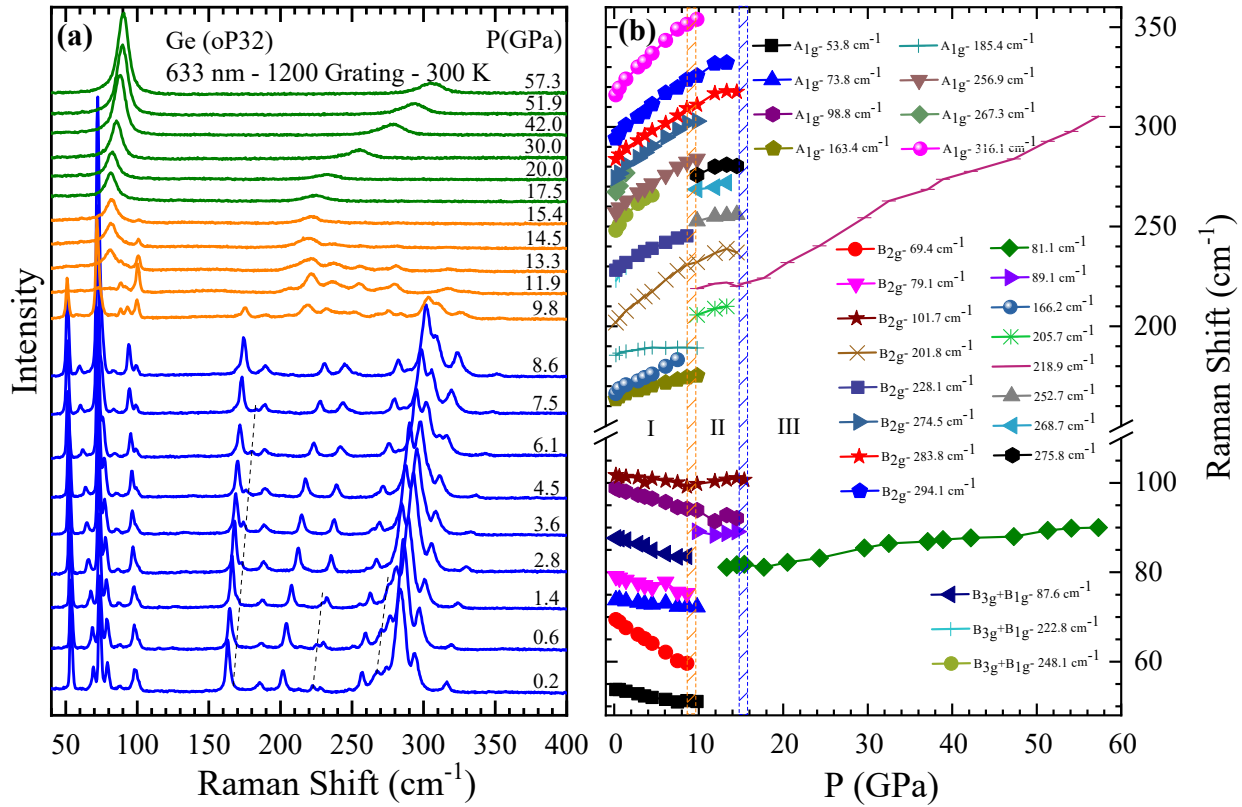

Fig. 2 (a) Room-temperature Raman spectra of Ge (oP32) over the frequency range of 40–400 cm$^{-1}$ for several selected pressures up to 57.3 GPa. Three phases are distinguished using different colors as follows: blue, orthorhombic; orange, intermediate structure; and green, β-tin phase. Dashed lines are guides for the eye and indicate the following three specific frequency changes: the peak at 166.2 cm$^{-1}$ appears at 0.2 GPa and is sustained up to 7.5 GPa, and the peaks corresponding to the $B_{3g} + B_{1g}$ mode at 222.8 cm$^{-1}$ and the $A_{1g}$ mode at 267.3 cm$^{-1}$ both merge with their respective adjacent peaks above 1.4 GPa. (b) Pressure-dependent Raman frequency shifts of several selected vibrational modes for a Ge (oP32) single crystal. Depending on the frequency as a function of pressure, three phases (I, II, and III) can be clearly distinguished, as indicated by the vertical dashed lines.

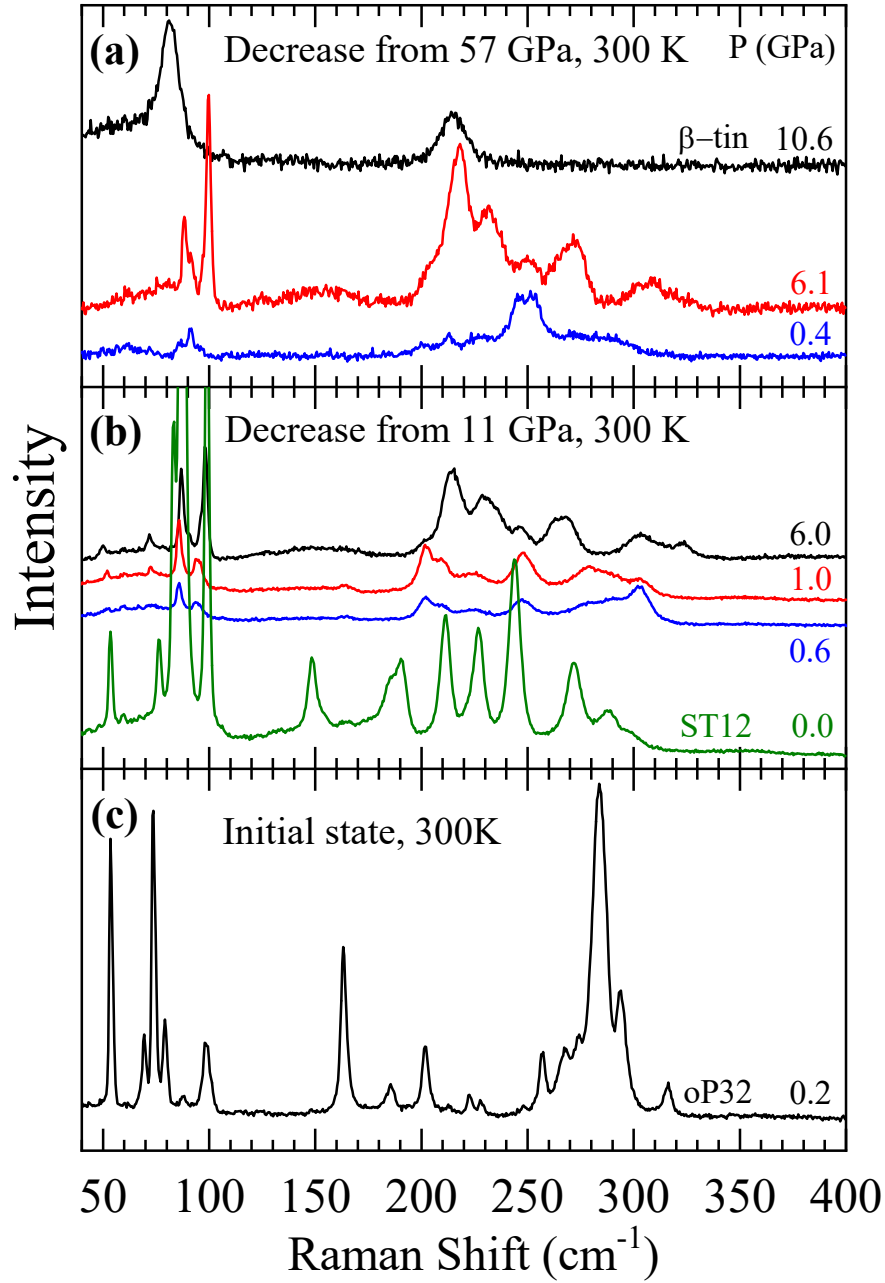

Fig. 3. Raman spectra obtained during decompression *vs.* the initial Raman spectrum. Room-temperature spectra were obtained during decompression from (a) 57 GPa and (b) 11 GPa. (c) Initial room-temperature Raman spectrum at 0.2 GPa.

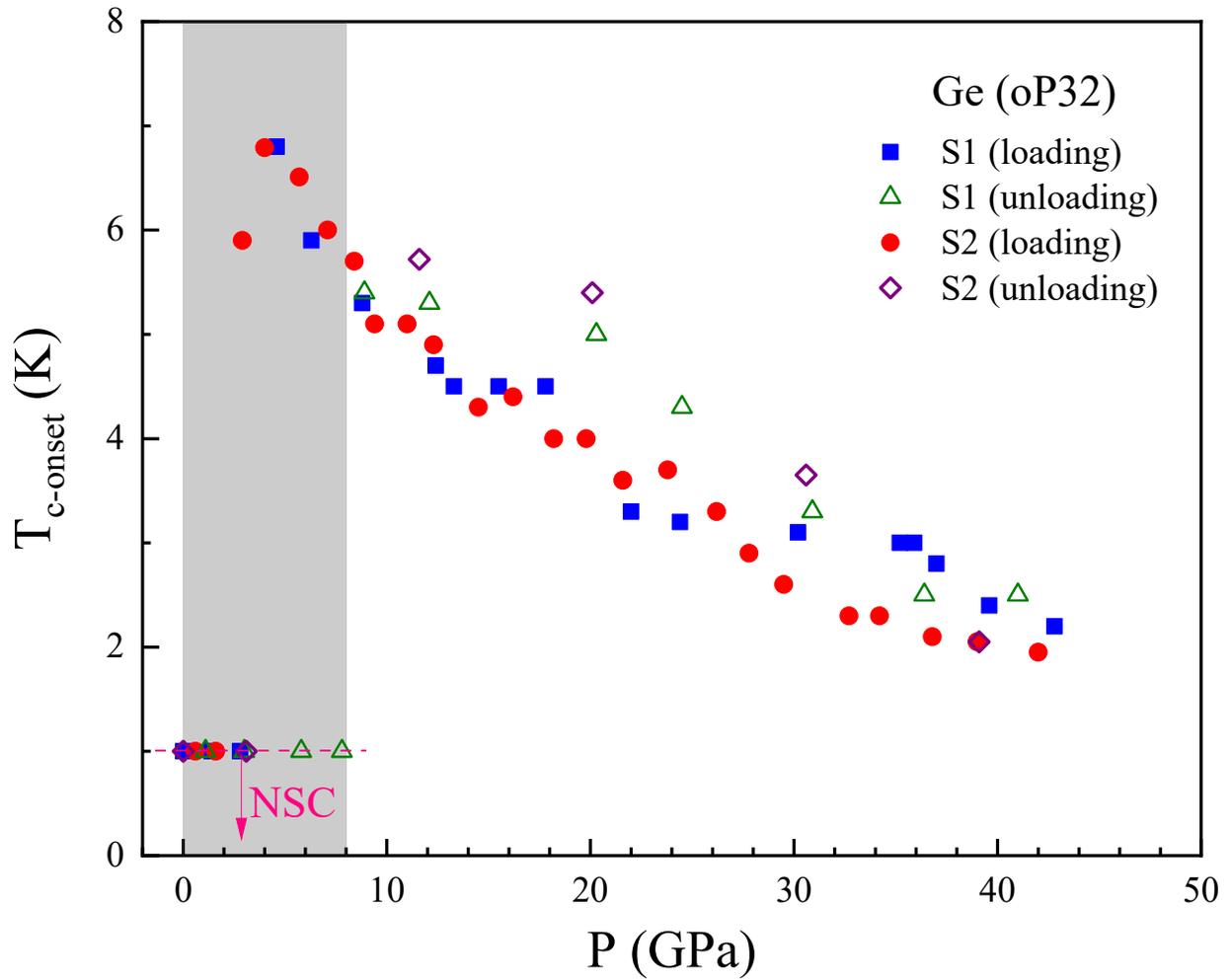

Fig. 4. $T_c$ as a function of pressure for single-crystal Ge (oP32). Solid symbols represent the results obtained during compression and the open symbols represent those obtained during decompression. No superconductivity signal was detected down to 1.9 K for the data along the pink dashed line. S1 and S2 represent two single crystals synthesized under the same conditions.

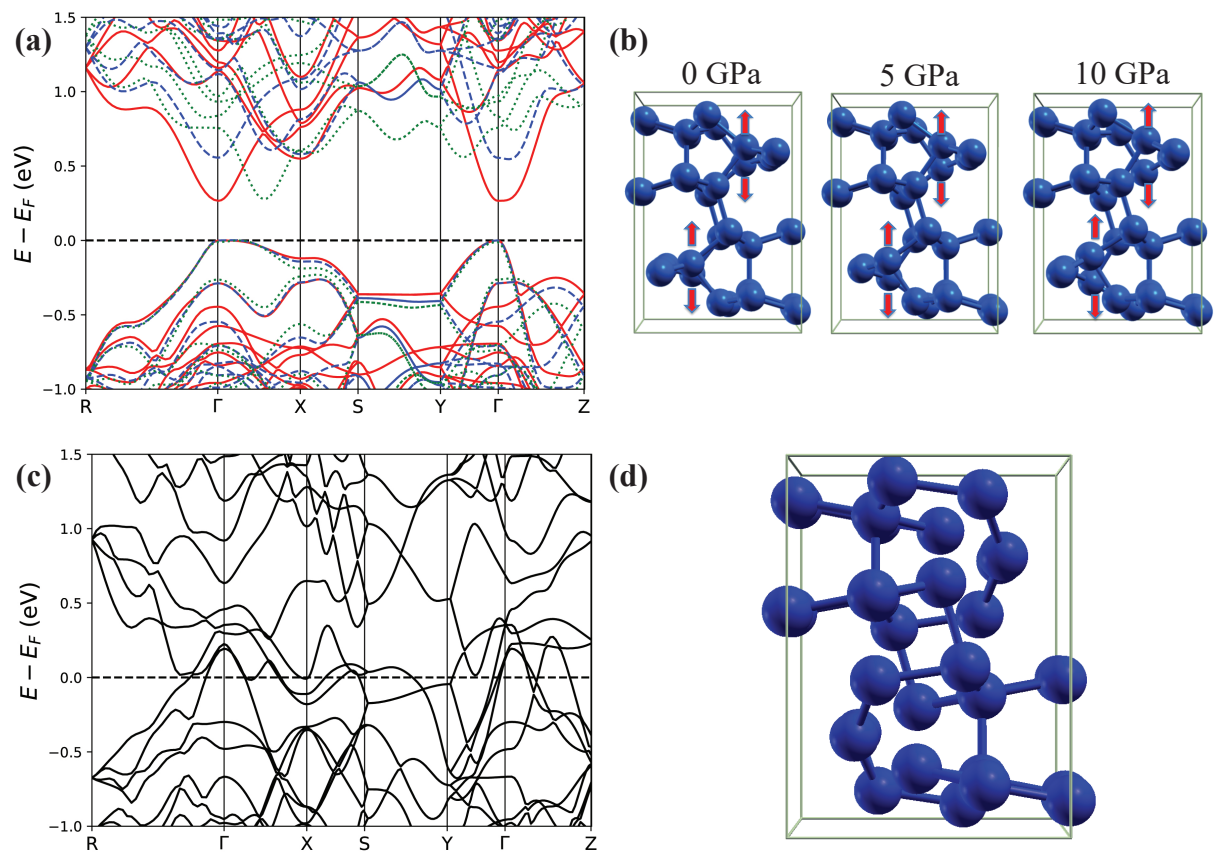

Fig. 5. (a) Band structure of Ge (oP32) under ambient pressure (red solid line) and under 5 GPa (blue dashed line) and 10 GPa (green dotted line). The horizontal dashed line indicates the valence band maximum. (b) Calculated crystal structure of Ge (oP32) at ambient and under 5 and 10 GPa. The arrows indicate displacement of atoms due to the applied pressure. (c) Band structure of Ge (oP32) under pressure of 12.5 GPa. The horizontal dashed line indicates the Fermi energy. (d) Calculated crystal structure of Ge (oP32) under 12.5 GPa.

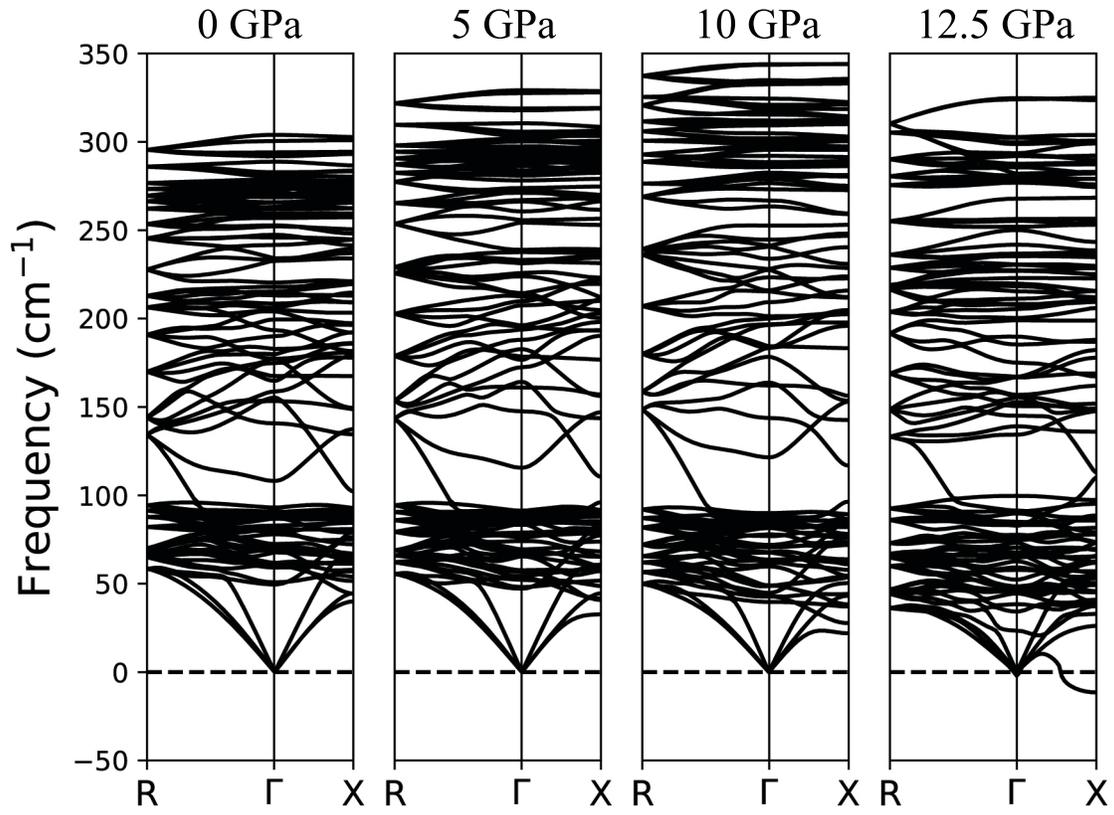

Fig. 6. Phonon dispersion relation of Ge (oP32) under various pressures up to 12.5 GPa.

Supplementary Information for:

# Pressure-induced superconductivity in a novel germanium allotrope

Liangzi Deng[1,#,*], Jianbo Zhang[2,#], Yuki Sakai[3], Zhongjia Tang[4], Moein Adnani[1], Rabin Dahal[1], Alexander P. Litvinchuk[1], James R. Chelikowsky[3,5], Marvin L. Cohen[6,7], Russell J. Hemley[8], Arnold Guloy[4], Yang Ding[2], Ching-Wu Chu[1,7,*]

#: These authors contributed equally to this work

* Corresponding: ldeng2@central.uh.edu, cwchu@uh.edu

1. Texas Center for Superconductivity and Department of Physics, University of Houston, Houston, Texas 77204, USA

2. Center for High Pressure Science and Technology Advanced Research, Beijing 100094, China

3. Center for Computational Materials, Oden Institute of Computational Engineering and Sciences, University of Texas at Austin, Austin, Texas 78712, USA

4. Texas Center for Superconductivity and Department of Chemistry, University of Houston, Houston, Texas 77204, USA

5. Department of Physics and McKetta Department of Chemical Engineering, University of Texas at Austin, Austin, Texas 78712, USA

6. Department of Physics, University of California at Berkeley, Berkeley, California 94720, USA
7. Materials Sciences Division, Lawrence Berkeley National Laboratory, Berkeley, California 94720, USA

8. Departments of Physics, Chemistry, and Earth and Environmental Sciences, University of Illinois Chicago, Chicago, Illinois 60607, USA

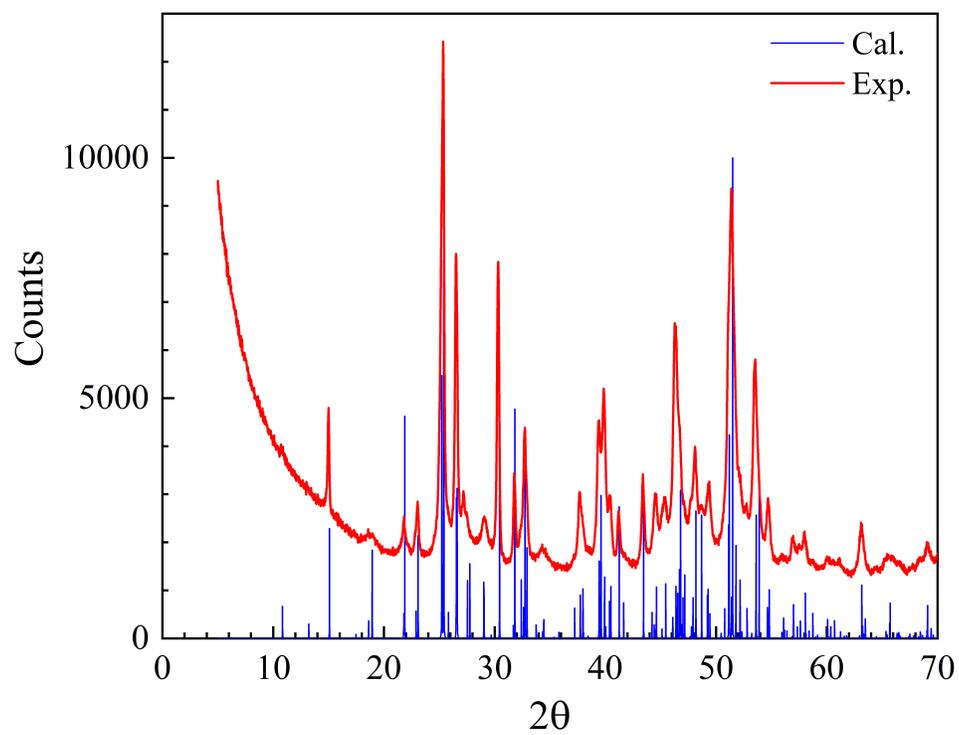

Fig. S1. Measured (red) and calculated (blue) powder XRD patterns for Ge (oP32). The calculated XRD pattern is based on the Ge (oP32) single-crystal structure determination results.

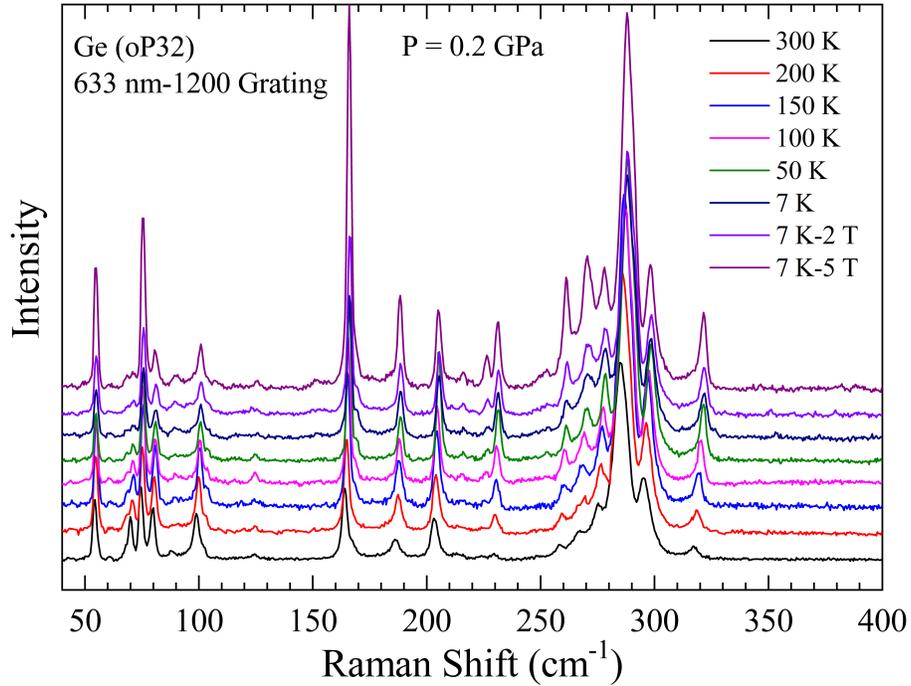

Fig. S2. Raman spectra of Ge (oP32) under 5 T in the frequency range of 40–400 cm$^{-1}$ from 300 to 7 K.

From the experimental Raman spectra of Ge (oP32) crystals at varying temperature shown in Fig. S1, the observed phonon lines become increasingly sharp with decreasing temperature. Indeed, the typical line width of the observed modes at room temperature is about 4.0 cm$^{-1}$ (2.5 cm$^{-1}$ at 7 K) in the high-frequency range and 2.0 cm$^{-1}$ (0.8 cm$^{-1}$ at 7 K) in the low-frequency range. This indicates unambiguously the high crystallinity of the Ge (oP32) allotrope. It has been reported that Ge (oP32) transforms into α-Ge at 636 K [7]. Our low-temperature and high-magnetic-field Raman measurements indicate that Ge (oP32) is a metastable allotrope and remains stable at temperatures ranging from 300 to 7 K and under magnetic fields up to 5 T. The effect of structural phase transition with varying temperature and magnetic field was excluded.

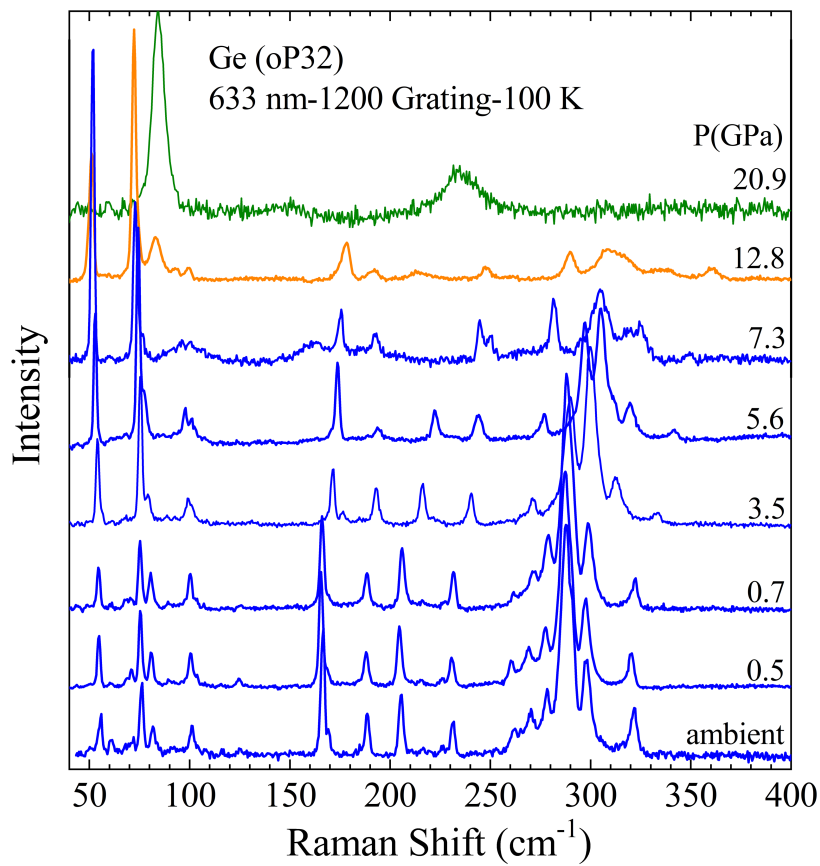

Fig. S3. Raman spectra of Ge (oP32) in the frequency range of 40–400 cm$^{-1}$ under pressure up to 20.9 GPa at 100 K.

Since the pressurization step is much more difficult to control at low temperatures than at room temperature, we did not undertake intensive acquisition of Raman spectra at low temperatures. However, by obtaining Raman spectra of Ge (oP32) under various pressures up to 20.9 GPa at 100 K, we were able to observe two pressure-induced phase transitions, distinguished by the three colors in Fig. S2. These results are consistent with the phase transitions observed at room temperature under high pressures.

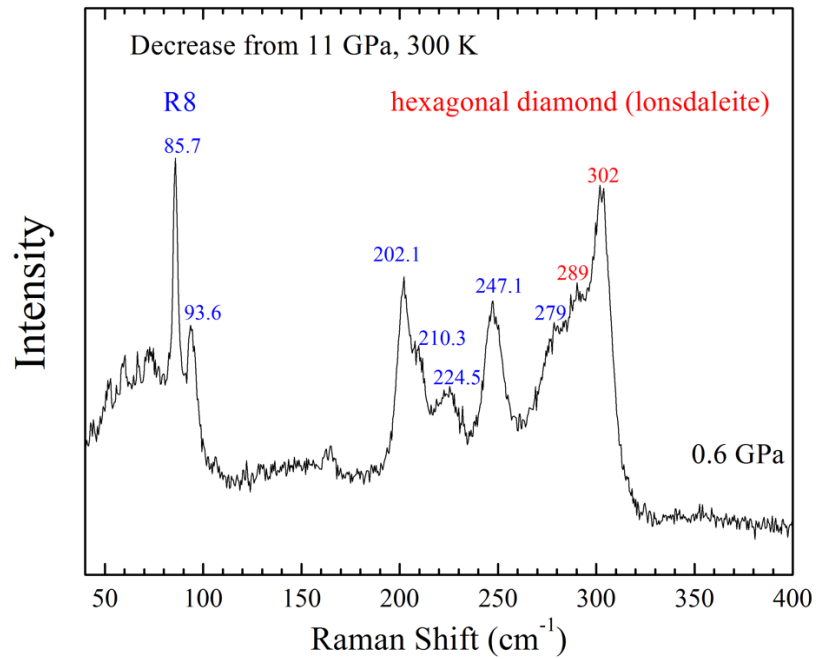

Fig. S4. Raman spectrum obtained from Fig. 3(b). Peaks at frequencies marked in blue are characteristic of the R8 phase and those in red are characteristic of the hexagonal diamond (lonsdaleite) phase.

Based on a previous report [24], 1) the Raman modes for the ST12 structure are 54, 75, 80, 85, 87, 88, 97, 152, 187, 194, 195, 216, 217, 226, 232, 249, 275, 276, 281, 284, and 295; 2) those for the BC8 structure are 87, 214, 227, 246, and 259; 3) those for the R8 structure are 83, 90, 203, 212, 223, 244, 247, and 278; and 4) those for the hexagonal diamond (lonsdaleite) Ge structure are 287, 301, and 302 (all frequencies in cm$^{-1}$ at 0 GPa). Based on a comparison with these results, we can confirm that at 0.6 GPa, the structure is mainly a mixture of the R8 and hexagonal diamond (lonsdaleite) phases.

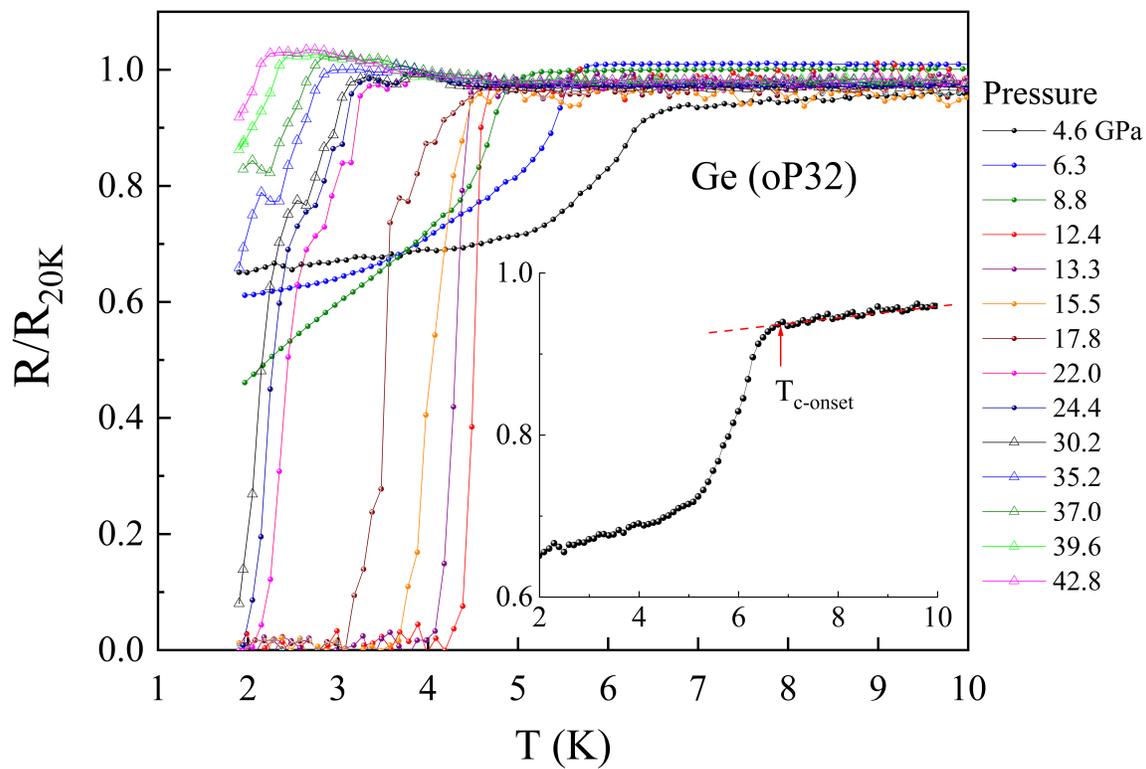

Fig. S5. Normalized resistance as a function of temperature for Ge (oP32) at different pressures up to 42.8 GPa during compression. Inset: approach used to determine $T_{c-onset}$.

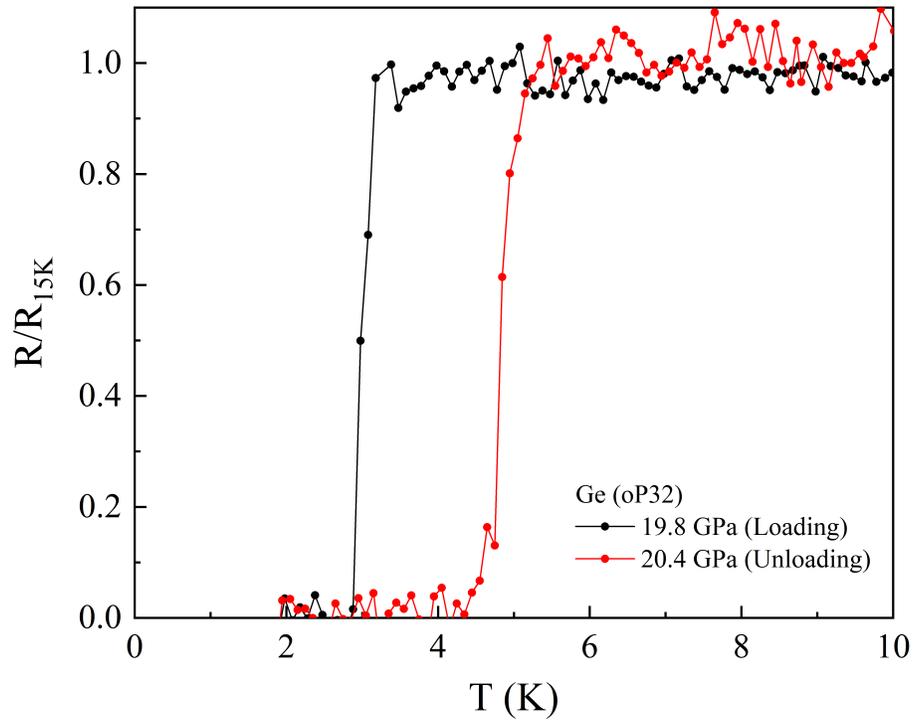

Fig. S6. Normalized resistance as a function of temperature for Ge (oP32) under 19.8 GPa (during compression) and 20.4 GPa (during decompression). A clear decompression-driven superconductivity enhancement is detected.

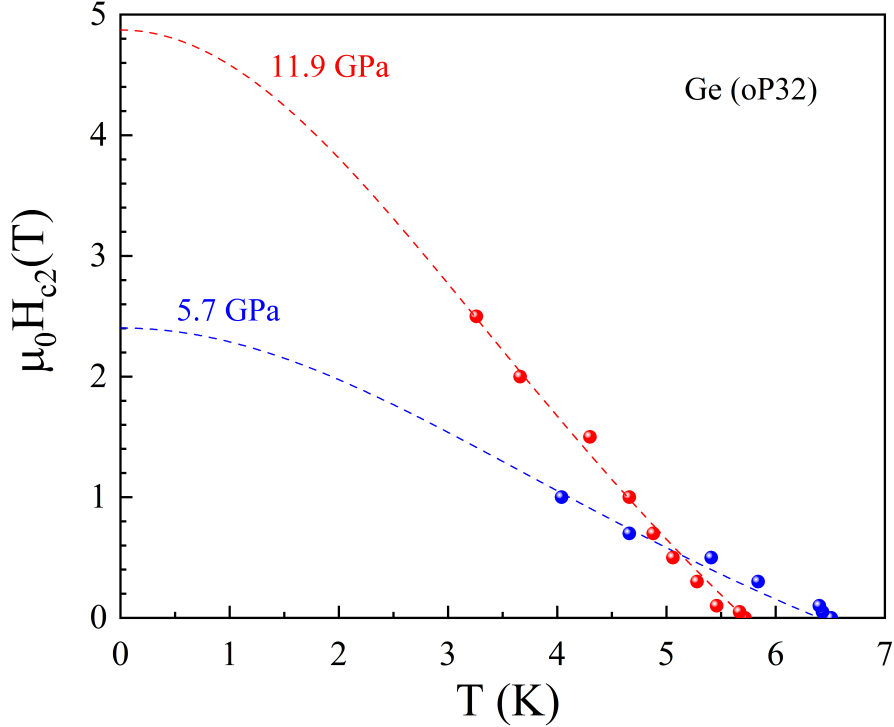

Fig. S7. H(T) phase diagram for Ge (oP32) under 5.7 GPa and 11.9 GPa. The dashed lines are fits of the results by the Ginzburg-Landau theory [26].

Table S1. Superconducting parameters [electron-phonon coupling strength ($\lambda$), logarithmic mean phonon frequency ($\omega_{\log}$), and superconducting transition temperature ($T_c$)] for the structures analyzed using density functional theory (DFT).

| Structure | $\lambda$ | $\omega_{\log}$ (cm$^{-1}$) | $T_c$ (K) |
|---|---|---|---|
| Under 12.5 GPa | 0.53 | 104 | 1.6 |
| After MD at 15 GPa | 1.23 | 100 | 9.2 |

Table S2. Crystal structures of Ge analyzed using DFT. The structure parameters are provided in the input file format for the Quantum ESPRESSO software package. For each structure, the CELL_PARAMETERS section represents cell vectors divided by alat and the ATOMIC_POSITIONS section shows the positions of atoms in angstroms.

**At 0 GPa**

```
CELL_PARAMETERS (alat = 15.51200000)
   1.000180431  -0.000004596   0.000000024
  -0.000006613   1.439147816   0.000002014
   0.000000023   0.000001331   0.952274040
```

ATOMIC_POSITIONS (angstroms)
| | | | |
|---|---|---|---|
| Ge | 4.667627216 | 0.977085917 | 1.954086432 |
| Ge | 3.543632175 | 10.844828439 | 5.862602587 |
| Ge | 3.543671596 | 6.883638777 | 1.954099671 |
| Ge | 4.667571407 | 4.938269501 | 5.862589090 |
| Ge | 2.725140913 | 2.579755035 | 1.954161713 |
| Ge | 5.486130360 | 9.242172256 | 5.862603929 |
| Ge | 5.486185432 | 8.486179692 | 1.954174757 |
| Ge | 2.725042806 | 3.335744512 | 5.862591444 |
| Ge | 3.994168073 | 10.464458960 | 1.954185357 |
| Ge | 4.217065921 | 1.357462154 | 5.862502738 |
| Ge | 4.217151860 | 4.557900380 | 1.954174184 |
| Ge | 3.994117646 | 7.264018382 | 5.862513507 |
| Ge | 8.018732937 | 4.884327628 | 0.674209772 |
| Ge | 0.192514665 | 6.937615301 | 7.142491515 |
| Ge | 0.192518891 | 6.937607713 | 4.582612205 |
| Ge | 8.018737244 | 4.884322312 | 3.234083197 |
| Ge | 0.192387272 | 10.790784443 | 3.234172239 |
| Ge | 8.018859380 | 1.031147536 | 4.582524426 |
| Ge | 8.018864888 | 1.031152061 | 7.142558016 |
| Ge | 0.192390711 | 10.790791869 | 0.674141096 |
| Ge | 1.272804236 | 2.957825260 | -0.000005776 |
| Ge | 6.938446508 | 8.864074994 | -0.000106526 |
| Ge | 6.938452436 | 8.864089047 | 3.908436870 |
| Ge | 1.272799963 | 2.957840612 | 3.908310686 |
| Ge | 5.745234873 | 1.741020607 | 7.716778849 |
| Ge | 2.466013790 | 10.080882999 | 0.099903378 |
| Ge | 2.465989431 | 10.080889760 | 3.808444620 |
| Ge | 5.745258357 | 1.741016208 | 4.008244256 |
| Ge | 2.466034303 | 7.647449015 | 4.008225759 |
| Ge | 5.745221062 | 4.174455197 | 3.808463390 |
| Ge | 5.745195434 | 4.174450238 | 0.099861379 |
| Ge | 2.466059686 | 7.647451390 | 7.716822117 |

**At 5 GPa**

CELL_PARAMETERS (alat = 15.09100000)
   1.000004753  -0.000000717   0.000000031
  -0.000001031   1.437996204  -0.000002052
   0.000000030  -0.000001370   0.960043794

ATOMIC_POSITIONS (angstroms)
| | | | |
|---|---|---|---|
| Ge | 4.624017734 | 0.887639019 | 1.916931454 |
| Ge | 3.363981636 | 10.604607552 | 5.750158539 |
| Ge | 3.364056037 | 6.629367617 | 1.916862233 |
| Ge | 4.624068391 | 4.862879037 | 5.750228722 |
| Ge | 2.642791185 | 2.363955287 | 1.916763385 |
| Ge | 5.345172858 | 9.128332915 | 5.750182857 |
| Ge | 5.345281323 | 8.105688478 | 1.916691339 |
| Ge | 2.642879097 | 3.386603308 | 5.750254401 |
| Ge | 3.954095294 | 10.102570185 | 1.916665257 |
| Ge | 4.033957465 | 1.389664086 | 5.750424805 |
| Ge | 4.033959231 | 4.360847274 | 1.916738249 |
| Ge | 3.954116706 | 7.131383694 | 5.750351591 |
| Ge | 7.823907986 | 4.760990134 | 0.682115007 |
| Ge | 0.164142715 | 6.731269427 | 6.984945916 |
| Ge | 0.164111062 | 6.731257825 | 4.515657525 |
| Ge | 7.823924955 | 4.761021064 | 3.151420954 |
| Ge | 0.164114053 | 10.502734863 | 3.151365793 |
| Ge | 7.823977090 | 0.989541598 | 4.515712233 |
| Ge | 7.823943719 | 0.989565917 | 6.985009108 |
| Ge | 0.164134089 | 10.502689736 | 0.682052359 |
| Ge | 1.252636570 | 2.875353955 | 0.000076821 |
| Ge | 6.735422190 | 8.617049222 | -0.000013657 |
| Ge | 6.735376333 | 8.617000074 | 3.833496410 |
| Ge | 1.252682856 | 2.875308872 | 3.833548775 |
| Ge | 5.596260750 | 1.681797086 | 7.543598639 |
| Ge | 2.391799133 | 9.810397838 | 0.123494694 |
| Ge | 2.391869510 | 9.810394855 | 3.709958515 |
| Ge | 5.596189602 | 1.681794242 | 3.957130154 |
| Ge | 2.391884252 | 7.423526624 | 3.957055324 |
| Ge | 5.596183516 | 4.068658275 | 3.710033966 |
| Ge | 5.596250476 | 4.068666225 | 0.123564026 |
| Ge | 2.391818886 | 7.423523477 | 7.543527679 |

**At 10 GPa**

CELL_PARAMETERS (alat = 14.70000000)
  0.999985576   0.000000064  -0.000000000
  0.000000092   1.433565380  -0.000005549
 -0.000000000  -0.000003771   0.974120611

ATOMIC_POSITIONS (angstroms)
| | | | |
|---|---|---|---|
| Ge | 4.628069366 | 0.783120009 | 1.894442191 |
| Ge | 3.152195018 | 10.376413033 | 5.683253658 |
| Ge | 3.152206967 | 6.358881201 | 1.894422865 |
| Ge | 4.628093968 | 4.800648372 | 5.683276055 |
| Ge | 2.545433451 | 2.068642882 | 1.894399144 |
| Ge | 5.234833887 | 9.090903830 | 5.683265826 |
| Ge | 5.234835439 | 7.644434693 | 1.894378504 |
| Ge | 2.545474752 | 3.515101937 | 5.683286886 |
| Ge | 3.961351994 | 9.700303324 | 1.894367787 |
| Ge | 3.818938813 | 1.459222787 | 5.683326750 |
| Ge | 3.818929497 | 4.124520404 | 1.894390109 |
| Ge | 3.961373980 | 7.035009324 | 5.683306876 |
| Ge | 7.640035113 | 4.634234685 | 0.697694169 |
| Ge | 0.140261102 | 6.525305025 | 6.879996903 |
| Ge | 0.140251129 | 6.525303742 | 4.486596930 |
| Ge | 7.640043792 | 4.634248604 | 3.091101589 |
| Ge | 0.140283092 | 10.210030614 | 3.091070643 |
| Ge | 7.640009792 | 0.949519324 | 4.486627467 |
| Ge | 7.639998369 | 0.949521648 | 6.880009142 |
| Ge | 0.140290886 | 10.210018081 | 0.697683205 |
| Ge | 1.251590625 | 2.791870152 | -0.000048561 |
| Ge | 6.528699946 | 8.367708284 | -0.000041344 |
| Ge | 6.528692975 | 8.367691948 | 3.788816175 |
| Ge | 1.251598516 | 2.791853975 | 3.788865076 |
| Ge | 5.439433624 | 1.619720875 | 7.414110503 |
| Ge | 2.340852398 | 9.539795818 | 0.163587499 |
| Ge | 2.340869126 | 9.539791701 | 3.625185240 |
| Ge | 5.439417219 | 1.619727892 | 3.952506528 |
| Ge | 2.340865442 | 7.195524272 | 3.952491128 |
| Ge | 5.439434729 | 3.963999999 | 3.625208282 |
| Ge | 5.439450465 | 3.964003853 | 0.163609106 |
| Ge | 2.340847862 | 7.195516057 | 7.414085940 |

**At 12.5 GPa**

CELL_PARAMETERS (alat = 14.39100000)
  1.000000000  0.000000000  0.000000000
  0.000000000  1.349340000  0.000000000
  0.000000000  0.000000000  0.998690000

ATOMIC_POSITIONS (angstroms)
| | | | |
|---|---|---|---|
| Ge | 4.811020319 | 0.565894469 | 1.901285183 |
| Ge | 2.805277324 | 9.717309233 | 5.703993258 |
| Ge | 2.805154436 | 5.703895337 | 1.901282461 |
| Ge | 4.811144585 | 4.579310813 | 5.703995761 |
| Ge | 2.328301288 | 0.937971313 | 1.901277333 |
| Ge | 5.287992802 | 9.345230485 | 5.703980886 |
| Ge | 5.287901757 | 6.075950961 | 1.901275429 |
| Ge | 2.328398212 | 4.207247996 | 5.703982037 |
| Ge | 4.144344623 | 8.518492303 | 1.901277808 |
| Ge | 3.471951546 | 1.764703090 | 5.703994723 |
| Ge | 3.471858544 | 3.380705424 | 1.901278107 |
| Ge | 4.144434403 | 6.902491717 | 5.703992526 |
| Ge | 7.635576891 | 4.184342281 | 0.715891835 |
| Ge | -0.019272581 | 6.098821095 | 6.889395156 |
| Ge | -0.019274281 | 6.098837593 | 4.518600007 |
| Ge | 7.635577847 | 4.184347195 | 3.086686360 |
| Ge | -0.019255315 | 9.322249283 | 3.086701859 |
| Ge | 7.635543516 | 0.960946837 | 4.518585795 |
| Ge | 7.635542203 | 0.960928579 | 6.889415657 |
| Ge | -0.019254860 | 9.322240979 | 0.715872045 |
| Ge | 1.692240680 | 2.572685821 | -0.000054490 |
| Ge | 5.924056056 | 7.710512926 | -0.000116022 |
| Ge | 5.924054977 | 7.710506785 | 3.802664868 |
| Ge | 1.692243832 | 2.572677931 | 3.802607000 |
| Ge | 5.319037460 | 1.395984310 | 7.329287626 |
| Ge | 2.297256803 | 8.887228167 | 0.275992300 |
| Ge | 2.297257839 | 8.887229163 | 3.526571262 |
| Ge | 5.319040248 | 1.395999606 | 4.078699440 |
| Ge | 2.297234599 | 6.533898143 | 4.078776674 |
| Ge | 5.319056615 | 3.749332011 | 3.526493258 |
| Ge | 5.319059153 | 3.749326800 | 0.276073120 |
| Ge | 2.297236801 | 6.533886999 | 7.329206746 |

**After molecular dynamics simulations at 15 GPa for 4 seconds**

CELL_PARAMETERS (alat = 14.08800000)
  0.955839037  0.013429069 -0.063489297
  0.024986635  1.390361492  0.001463685
 -0.066026185 -0.003764123  0.954763073

ATOMIC_POSITIONS (angstroms)
| | | | |
|---|---|---|---|
| Ge | 5.616123439 | 1.030231455 | 0.293017263 |
| Ge | 2.648981220 | 7.723179877 | 5.730778854 |
| Ge | 0.743109168 | 5.683774974 | 0.721547763 |
| Ge | 3.837037867 | 2.964374388 | 6.133946725 |
| Ge | 0.958285234 | 0.804395564 | 2.106617660 |
| Ge | 5.405293976 | 9.130901960 | 4.977528401 |
| Ge | 3.125137694 | 6.935163472 | 1.109107009 |
| Ge | 1.589101396 | 2.473274276 | 4.255547663 |
| Ge | 5.441398881 | 8.409822086 | 0.325639391 |
| Ge | 3.838272201 | 1.684118472 | 2.183936342 |
| Ge | 2.028149346 | 3.721871645 | 1.998699727 |
| Ge | 4.907429602 | 5.882384791 | 6.024623822 |
| Ge | 6.536138237 | 3.204102716 | 1.594395352 |
| Ge | -1.646929300 | 6.265513473 | 9.514307637 |
| Ge | 0.007971431 | 7.107902753 | 5.625886647 |
| Ge | 4.331812878 | 7.196518546 | 3.733264456 |
| Ge | 0.641263123 | 7.921851114 | 3.167771807 |
| Ge | 8.659641194 | -0.410320323 | 3.924033779 |
| Ge | 6.453512386 | 3.582576263 | 6.067112137 |

```
Ge     0.573763369  10.682765820  -0.898032968
Ge     1.988659442   2.137924730  -0.075162186
Ge     8.229325959   8.770195939   0.229483112
Ge     6.042114887   9.946908441   2.520134388
Ge    -0.185745265   4.727790002   4.283291235
Ge     3.699529183   0.814141273   4.693315844
Ge     3.420161306  10.144946788  -0.086804760
Ge     3.402697145   9.327974382   2.410183583
Ge     6.049699760   1.960335828   3.852028769
Ge     2.163943948   5.875665958   3.436198392
Ge     4.273352648   4.213852352   3.876116689
Ge     4.367505415   4.576826416   1.088347368
Ge     2.025921160   5.001705299   5.944841406
```